\documentclass{article}

\usepackage[preprint]{neurips_data_2022}

\usepackage[utf8]{inputenc} 
\usepackage[T1]{fontenc}    
\usepackage{hyperref}       
\usepackage{url}            
\usepackage{booktabs, cellspace, tabularx}       
\usepackage{amsfonts}       
\usepackage{nicefrac}       
\usepackage{amsmath}
\usepackage{microtype}      
\usepackage[table,RGB]{xcolor}         
\usepackage{times}
\usepackage{enumitem} 
\usepackage{appendix}
\usepackage{latexsym}
\usepackage{multirow}
\usepackage{color,soul}
\usepackage{multicol}
\usepackage{graphicx}
\usepackage{newfloat}
\usepackage{placeins}
\usepackage{float}
\usepackage{afterpage}
\usepackage{listings}

\lstdefinestyle{java}{language=Java, morekeywords={function,var}}
\lstdefinestyle{C++}{language=C++}
\lstdefinestyle{C}{language=C}
\lstdefinestyle{py}{language=Python}
\lstdefinestyle{Csharp}{language=[Sharp]C}
\lstdefinestyle{PHP}{language=PHP, morekeywords=function}
\newcommand{\dname}{XLCoST}
\newcolumntype{P}[1]{>{\centering\arraybackslash}p{#1}}

\title{XLCoST: A Benchmark Dataset for Cross-lingual Code Intelligence}

\author{Ming Zhu, Aneesh Jain, Karthik Suresh, Roshan Ravindran, Sindhu Tipirneni,\\ \textbf{Chandan K. Reddy} \\ Department of Computer Science, Virginia Tech, Arlington, VA\\
\texttt{\{mingzhu, aneeshj, karthiks, roshan14, tsaisindhura\}@vt.edu,} \\ \texttt{reddy@cs.vt.edu}}

\begin{document}

\maketitle

\begin{abstract}
Recent advances in machine learning have significantly improved the understanding of source code data and achieved good performance on a number of downstream tasks. Open source repositories like GitHub enable this process with rich unlabeled code data. However, the lack of high quality labeled data has largely hindered the progress of several code related tasks, such as program translation, summarization, synthesis, and code search. This paper introduces $\dname$, \textbf{Cross-L}ingual \textbf{Co}de \textbf{S}nippe\textbf{T} dataset, a new benchmark dataset for cross-lingual code intelligence. Our dataset contains fine-grained parallel data from 8 languages (7 commonly used programming languages and English), and supports 10 cross-lingual code tasks. To the best of our knowledge, it is the largest parallel dataset for source code both in terms of size and the number of languages. We also provide the performance of several state-of-the-art baseline models for each task. We believe this new dataset can be a valuable asset for the research community and facilitate the development and validation of new methods for cross-lingual code intelligence\footnote{\url{https://github.com/reddy-lab-code-research/XLCoST}}. 
\end{abstract}

\section{Introduction}

Recent advances in machine learning have benefited a number of code related tasks, such as code translation, code summarization, and code synthesis. Open-source code repository websites like Github provide enormous amount of source code data, which enables the training of large-scale programming language models such as CodeBERT \citep{feng2020codebert}, PLBART \citep{ahmad2021unified}, TransCoder \citep{roziere2020unsupervised} and CodeT5 \citep{wang2021codet5}. These extensively pre-trained models have shown superior performance on benchmark datasets like CodeXGLUE \citep{lu2021codexglue}.

Although open-source code data is abundant in quantity, it has several disadvantages when being used as training data for code-related models. First, most of the available code data is unlabeled. For tasks like Code Translation, Code Summarization, and Code Synthesis, high quality parallel data is critical for model training. However, it is difficult to mine parallel data from open-source projects. Second, labeled data is usually small in size. For example, the code translation data introduced in \cite{zhu2022multilingual} only has around 70 programs for testing and 50 programs for validation. Due to the small size of evaluation data, the models trained on this dataset may not be thoroughly evaluated. Moreover, the available labeled datasets usually only cover a limited number of languages. For example, the Code Translation dataset in CodeXGLUE only covers 2 languages, Java and C\#. Because of the scarcity of labeled data in some programming languages, code tasks in some low-resource languages remain unexplored. 

In this paper, we introduce $\dname$, a machine learning benchmark dataset that contains fine-grained parallel data in 7 commonly used programming languages (C++, Java, Python, C\#, Javascript, PHP, C), and natural language (English). The data is parallel across 7 languages, at both code snippet level and program level. This means that, given a program in one language, the dataset contains the same program in up to 6 other programming languages. Each program is divided into several code snippets, and programs in all the languages are aligned at the snippet level. 
Moreover, each of the snippets is accompanied with a comment, and  the comment for a particular snippet is the same across all the languages. Table \ref{tab:dataset_survey} presents a comparative analysis of $\dname$ in terms of the number of available parallel data samples against other widely used parallel code datasets.
The dataset contains around 1 million parallel snippets and 123K parallel programs in total, which is significantly larger than many available parallel code datasets. We believe that this dataset is a valuable asset for the research community and can potentially benefit a number of code-related research problems.

To further facilitate the development and evaluation of models with a focus on source code, we also introduce 10 different cross-lingual tasks. These tasks can be divided into two categories: Generation and Retrieval. The generation tasks include Code Translation (Code-to-Code), Code Summarization (Code-to-Text), and Code Synthesis (Text-to-Code); the retrieval tasks include NL (Natural Language) Code Search and XL (Cross-Lingual) Code Search. Each task is at both snippet and program level. 

To evaluate how challenging the tasks are with the proposed dataset, we run experiments on all the 10 tasks with a number of state-of-the-art baseline models. We also conduct an empirical study to understand how the model design relates with the performance on different tasks with $\dname$ dataset. The primary contributions of this paper are as follows:
\begin{itemize} [leftmargin=*]
    \item We introduce a new dataset which is parallel across 8 languages (7 programming languages and English) at both snippet level and program level. To the best of our knowledge, it is the largest \textbf{parallel} dataset for source code in both size and number of languages.
    \item We formulate 10 different cross-lingual tasks to facilitate the development and evaluation of models in this domain.
    \item We run experiments for all the 10 tasks on the proposed dataset with a number of state-of-the-art baseline models and provide insights about model design for the new challenges.
\end{itemize}






\begin{table*}[]
\fontsize{8.15pt}{8.15pt}\selectfont
\rmfamily
 \aboverulesep=0ex
 \belowrulesep=0ex
  \renewcommand{\arraystretch}{1.3}
\setlength{\tabcolsep}{1.2pt}

\caption{\label{tab:dataset_survey} Comparison against other parallel code datasets (Py - Python, JS - JavaScript). Column "Size" refers to the number of parallel data pairs. *This number is for single programs, not pairs.}

\begin{tabular}{p{1.6cm}p{1cm}p{1.35cm}p{3.55cm}p{1.25cm}p{4.45cm}}
\toprule
\multicolumn{1}{l}{\textbf{Dataset}} & \multicolumn{1}{l}{\textbf{Alignment}}  & \multicolumn{1}{l}{\textbf{Task}}  & \multicolumn{1}{l}{\textbf{Labelling}}  & \multicolumn{1}{l}{\textbf{Size}}  & \multicolumn{1}{l}{\textbf{Languages}}  \\ \midrule
CodeNet & Program & Multiple & Solutions to the same problem & 13.9M\textsuperscript{\rm *} & 55 programming languages \\ 
AVATAR & Program & Translation & Solutions to the same problem & 57,414 & Java, Py \\
CodeXGLUE & Method & Multiple & Matching function names & 11,800 & Java, C\# \\
CoST & Snippet & Translation & Matching code comments & 132,046 & C++, Java, Py, C\#, JS, PHP, C \\ \midrule
XLCoST & Snippet & Multiple & Matching code comments & 1,002,296 & C++, Java, Py, C\#, JS, PHP, C, English\\

 \bottomrule
\end{tabular}

\end{table*}

\section{Related work}

\textbf{Parallel Code Data} CodeXGLUE \citep{lu2021codexglue} is a popular benchmark that includes 14 datasets for 10 code related tasks. The tasks include clone detection, code translation, natural language code search, etc. However, this benchmark does not contain datasets with parallel codes from more than 2 languages. CoST \citet{zhu2022multilingual} is a code translation dataset for 7 programming languages. However, it is relatively small and only supports translation task. AVATAR \citep{ahmad-etal-2021-avatar} presents another parallel dataset for Java-Python translation. The authors collect multiple solutions for problems scraped from competitive programming websites and then form $n^2$ possible combinations of parallel data. This is also constrained to only 2 languages. Project CodeNet \citep{puri2021project} has an abundance of parallel programs in a wide range of languages. However, the programs are significantly different in logic and structure, thus the alignment is of low quality.

\textbf{Cross-Lingual Code Tasks} Several tasks in the code domain are related to our work, including Code Translation, Code Summarization, Code Synthesis, and Code Search. CodeBERT \citep{feng2020codebert} pre-trained a BERT \citep{kenton2019bert} based encoder on the source code, and then added a decoder to perform end-to-end training on code translation. CodeBERT is also used for Code Search tasks. PLBART \citep{ahmad2021unified} utilized an existing natural language translation model, BART \citep{lewis2020bart}, and also pre-trained it with source code. CodeTransformer \citep{zuegner_code_transformer_2021} uses language agnostic features computed from the source code and its abstract syntax tree for code summarization. 
OpenAI's Codex \citep{chen2021evaluating} framework makes use of GPT \citep{radford2018improving} language models fine-tuned on publicly available code from GitHub for code related downstream tasks. However, most of the models only explored a limited number of languages, due to the scarcity of multilingual parallel data.

\section{The $\dname$ dataset}
The data for $\dname$ was collected from GeeksForGeeks\footnote{https://www.geeksforgeeks.org/}, which is a website that houses thousands of data structures and algorithm problems along with solutions in up to 7 different programming languages - C++, Java, Python, C\#, Javascript, PHP, and C. 
According to GeeksForGeeks, the solution programs for the same problem follow the same structure, down to the variable names. This results in the programs being semantically consistent across the different languages. In most cases, the programs for the same problem share the same set of comments in the same order, which indicates that they are parallel to the snippet level. This is where the fine-grained alignment in $\dname$ comes from.

\begin{figure*}[ht]
    \centering
    \includegraphics[trim={0.1cm 0.1cm 0.1cm 0.1cm},clip, width=1.0\linewidth]{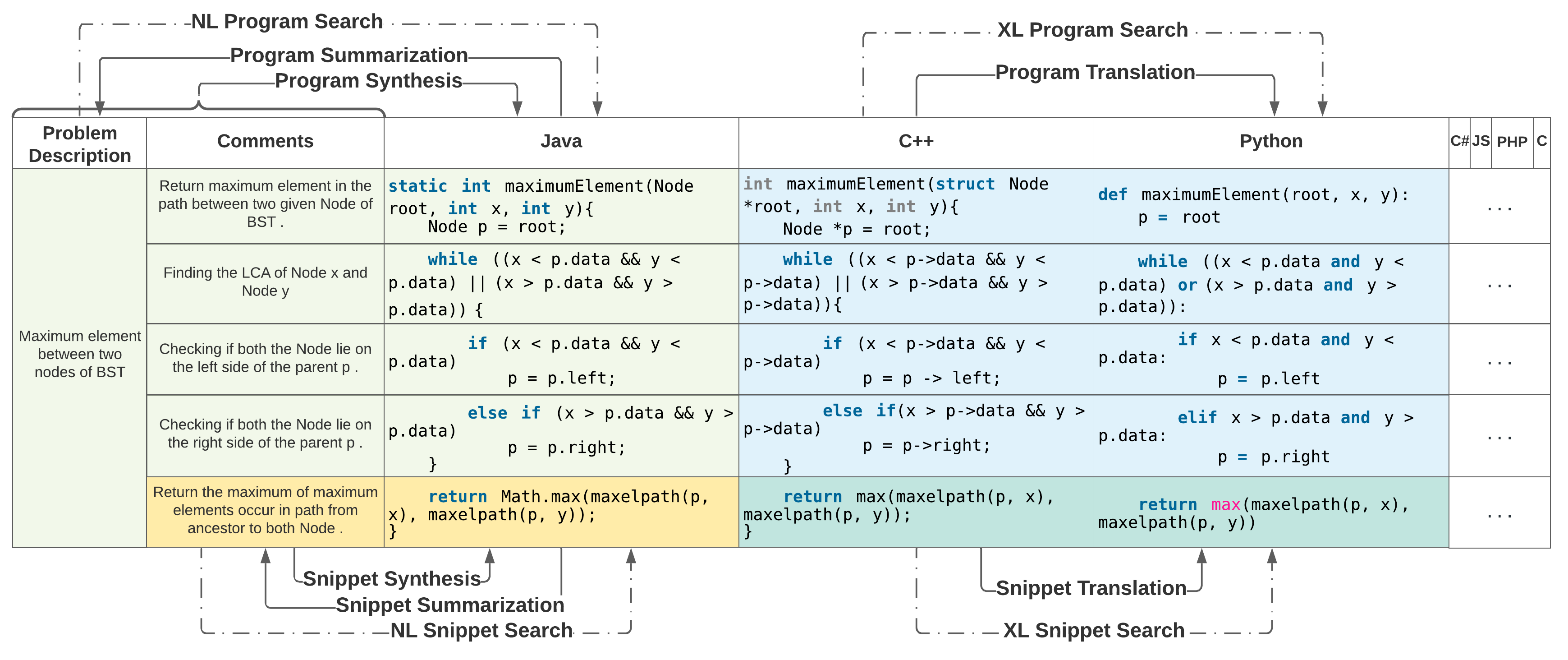}
    \caption{An illustration of the data and the tasks. 
The first column is the Problem Description; each cell in the second column is a Comment; each cell from the third column is a code Snippet. The combination of all the code snippets in a column is a Program (truncated due to space limitation). The arrows show the input and output data for each task. Solid lines are for generation tasks and dashed lines are for retrieval tasks. Note that the Program Synthesis task uses both Problem Description and Comments as input.  } 
    \label{fig:data_task}
\end{figure*}

\subsection{Definitions}
\textbf{Problems}: The problems are mostly about data structures and algorithms, as they are mainly designed for tutoring and coding interview preparation. Each problem has programs as solutions in up to 7 programming languages. \\
\textbf{Programs}: A program is a solution to a problem in a specific programming language. Each problem in this dataset may contain up to 7 programs (one for each language). The programs for the same problem share similar logic and structure.\\
\textbf{Snippets}: The code between two consecutive comments in a program is termed as a snippet (code before the first comment and after the last comment are also included). On an average, each program contains 8.81 snippets.  \\
\textbf{Description}: Each problem also has a short description, for example, ``Maximum Consecutive Increasing Path Length in Binary Tree." \\
\noindent \textbf{Comments}: The comments in each program in this dataset. The programs are well commented and each program has an average of around 9 comments.

\subsection{Data Characteristics}
The final dataset consists of 11,265 programming problems. As shown in Table \ref{tab:data_stats_small}, there are 
57,661 unique programs. Each program consists of 8.81 snippets on average, which results in 
509,091 snippets. A detailed statistics table for the translation task is available in Appendix A.2.

\textbf{Multilingual}: The dataset contains parallel data in 8 languages (7 commonly used programming languages and English). \\
\textbf{Parallel}: The dataset contains 4 types of parallel data, snippet-to-snippet,  program-to-program, snippet-to-comment, program-to-problem (and comments) which further enables 10 different tasks.\\
\textbf{Finely-aligned}: The data is parallel at both snippet level and program level. To the best of our knowledge, this dataset is the finest-aligned among parallel code datasets.\\
\textbf{Large}: It is the largest parallel dataset for source code in terms of both size and number of languages.\\
\textbf{Simple}: Each program in this dataset is standalone without dependency on other programs. It ensures that the complexity of the tasks is controllable. 

\begin{table*}[]
 \aboverulesep=0ex
 \belowrulesep=0ex
\setlength{\tabcolsep}{3pt}
\renewcommand{\arraystretch}{1.3}
\scriptsize
\rmfamily

\caption{\label{tab:data_stats_small} The train-valid-test split and basic statistics of $\dname$ data. SN - Snippets; PR - Program.}

\begin{tabular}{ p{1.35cm}|P{0.55cm}@{\hspace{8pt}}P{0.65cm}P{0.6cm}P{0.6cm}P{0.6cm}P{0.55cm}P{0.5cm}P{0.48cm} @{\hspace{8pt}}|P{0.52cm}P{0.52cm}P{0.52cm}P{0.52cm}P{0.52cm}P{0.52cm}P{0.52cm}P{0.5cm}}

\toprule\multicolumn{1}{ c|}{}             & \multicolumn{8}{c|}{\textbf{Snippet-level}}                                                                            & \multicolumn{8}{c }{\textbf{Program-level}}                                                                            \\ \midrule 
\textbf{Split} & \textbf{C++}   & \textbf{Java}  & \textbf{Py} & \textbf{C\#}   & \textbf{JS}    & \textbf{PHP}   & \textbf{C} & \textbf{Total}     & \textbf{C++}   & \textbf{Java}  & \textbf{Py} & \textbf{C\#}   & \textbf{JS}    & \textbf{PHP}   & \textbf{C} & \textbf{Total}      \\\midrule
train & 93847 & 91089 & 81207 & 87583 & 70649 & 18027 & 3763 & 446165 & 9797 & 9623 & 9263 & 9345 & 8590 & 3087 & 463 & 50168 \\
valid & 4432 & 4460 & 3946 & 4436 & 3829 & 930 & 350 & 22383 & 492 & 494 & 472 & 491 & 475 & 158 & 60 & 2642 \\
test & 8118 & 8154 & 7293 & 8013 & 7033 & 1682 & 250 & 40543 & 909 & 911 & 887 & 899 & 886 & 308 & 51 & 4851 \\
\arrayrulecolor{black!50}\bottomrule
total & 106397 & 103703 & 92446 & 100032 & 81511 & 20639 & 4363 & 509091 & 11198 & 11028 & 10622 & 10735 & 9951 & 3553 & 574 & 57661\\\arrayrulecolor{black}
 
               \hline
     \multicolumn{1}{c|}{}                                                                           & \multicolumn{8}{c|}{} & \multicolumn{8}{c}{}\\[-2.7ex]
     \hline
\textbf{Stats} & \textbf{C++}   & \textbf{Java}  & \textbf{Py} & \textbf{C\#}   & \textbf{JS}    & \textbf{PHP}   & \textbf{C} & \textbf{Avg.}     & \textbf{C++}   & \textbf{Java}  & \textbf{Py} & \textbf{C\#}   & \textbf{JS}    & \textbf{PHP}   & \textbf{C} & \textbf{Avg.}      \\\midrule
\# lines/code & 3.41 & 3.71 & 2.41 & 3.82 & 3.23 & 4 & 4.05 & 3.37 & 32.45 & 34.93 & 20.54 & 35.64 & 26.47 & 23.23 & 31.5 & 29.71 \\
\# tokens/code & 21.52 & 24.1 & 21.63 & 23.06 & 22.52 & 28.14 & 25.37 & 22.83 & 205 & 227.1 & 188.5 & 215.3 & 184.6 & 163.5 & 198 & 202 \\
\# tokens/text & 8.25 & 8.14 & 7.97 & 8.23 & 7.96 & 8.45 & 9.67 & 8.15 & 10.68 & 10.67 & 10.75 & 10.7 & 10.87 & 9.91 & 8.19 & 10.66 \\
\# SN/PR & \textendash & \textendash & \textendash & \textendash & \textendash & \textendash & \textendash & \textendash & 9.52 & 9.42 & 8.51 & 9.33 & 8.2 & 5.81 & 7.77 & 8.81\\
 \arrayrulecolor{black}\bottomrule
\end{tabular}

\end{table*}

\subsection{Data Collection and Processing}
The data was scraped from different sub-pages of the GeeksForGeeks website. A majority of the problems on this site fall under two categories - Data Structures and Algorithms. More details are included in Appendix A.3. The IP policies and regulations for GeeksForGeeks were carefully followed and we confirm that no data privacy policy was violated when collecting the data. 

After collecting the data, we first removed duplicate problems, as some problems might be presented in multiple subcategories. Then we extracted problem description and solution programs in each available language from the page. Each program was sliced into code snippets by splitting at the comments, after which the comments and docstrings were removed from the programs. Any personal information such as the name of the code's contributor, was also removed from both the comments and the codes at this time. Eventually, we get 4 types of information from one page: 1) Problem Description; 2) Parallel programs in different languages; 3) Code Snippets; 4) Code Comments. 

\subsubsection{Data Alignment}
The snippet-level alignment was done by matching comments in the solution programs (for the same problem) across different languages. As mentioned earlier, GeeksForGeeks programs follow a standard template, because of which the comments in different language programs (for the same problem) align parallelly in most cases. This yields parallel snippets that have the same functionality across different languages.

\textbf{Misalignment detection:}
In some cases, the comments in different solution programs are not aligned. The misalignment can come from different numbers of comments, and the differences in the comment content. This is usually due to some solution program not strictly following the guidelines and templates. 
For solution programs with the same number of comments, we evaluate the alignment by calculating the average similarity score of each pair of comments in the two programs (using Python \textit{difflib.SequenceMatcher}\footnote{https://docs.python.org/3/library/difflib.html})). If the average score is below a certain threshold (80\% in our case), it would be categorized as misalignment and manual checking would be needed. Solution programs with different number of comments were automatically categorized as misaligned and sent for manual checking.

\textbf{Manual checking and aligning:}
Manual checking was performed by two of the authors with good knowledge of programming languages and their functionalities. Based on the differences in number of comments, the misaligned programs were split into the following categories:\\
\underline{Category 0}: The programs have the same number of comments. The type of misalignment usually only is due to different wording in the comments and can be easily fixed. \\
\underline{Category $k$}: The difference in number of comments is $k$.  When $k<3$, extra comments needed to be discarded in some cases and code from these comments was moved to appropriate snippets to preserve the alignment with other languages. In some cases, there were also missing comments which had to be added along with the moving of the appropriate code block as in the previous case. When $k>=3$, the programs will be discarded.

\subsubsection{Data Splitting}
Since the parallel programs are within each problem, splitting the data at problem level can naturally avoid data leakage. However, during the data processing, we noticed that some problems are very similar. For example, "Check if a large number is divisible by 3 or not" and "Check whether a large number is divisible by 53 or not". If one problem goes to the training set and the other goes to the test set, it can lead to potential data leakage and bias. To address this concern, we first clustered all the similar problems into groups, and make the split at the group-level. In this way, we can ensure that similar problems go to the same split.
To do so, we first calculate the similarity score (using Python \textit{difflib.SequenceMatcher}) between every two pairs of problem descriptions, and group all the problems using various similarity score thresholds (60\%-80\%) based on length of the descriptions. The final split ratio in the data is around 85-5-10 for train-validation-test sets. The detailed steps for data splitting are included in Appendix A.4.

\begin{table*}[]
\centering
 \scriptsize
\rmfamily
 \aboverulesep=0ex
 \belowrulesep=0ex
  \renewcommand{\arraystretch}{1.6}
\setlength{\tabcolsep}{1.4pt}

\caption{\label{tab:task_overview}An overview of the tasks. All the tasks have pairwise data at both snippet-level and program-level in 7 programming languages, C++, Java, Python, C\#, Javascript, PHP, and C. The tasks can be divided into two categories, generation and retrieval. The generation tasks include Code Translation, Code Summarization and Code Syntheis; the retrieval tasks include NL (natural language) Code Search and XL (Cross-Lingual) Code Search. 
}

\begin{tabular}{@{}l@{\hspace{5.2pt}}p{1.95cm}lp{1.6cm}@{\hspace{1pt}}p{6cm}}
\toprule

\multicolumn{1}{l}{\textbf{Category}}     & \multicolumn{1}{l}{}                              & \multicolumn{1}{l}{\textbf{Task}}                     & \multicolumn{1}{l}{\textbf{Data}}                                              & \multicolumn{1}{l}{\textbf{Description}}
\\ \midrule
\multirow{6}{*}{Generation} &  \multirow{2}{=}{Code Translation (Code-to-Code)}   & Snippet Translation       & 872K/47K/83K  & Translate code snippet across programming languages  
\\ \cmidrule{3-5} 
                            &                                 & Program Translation       & 106K/6K/11K                           & Translate program across programming languages                           
                            \\ \cmidrule{2-5}
                            &  \multirow{2}{=}{\fontsize{6.5pt}{6.5pt}\selectfont Code Summarization \scriptsize (Code-to-Text)}   & Snippet Summarization     & 446K/22K/41K                            & Generate comment for given code snippet                          
                            \\ \cmidrule{3-5}
                            &                                 & Program Summarization     & 50K/3K/5K                             & Generate problem description for given program                
                            \\ \cmidrule{2-5}
                            &  \multirow{2}{=}{Code Synthesis (Text-to-Code)}   & Snippet Synthesis         & 446K/22K/41K                              & Generate code snippet giving comment               
                            \\ \cmidrule{3-5}
                            &                                 & Program Synthesis         & 50K/3K/5K       & Generate program giving problem description and comments                                   
                            \\ \cmidrule{1-5}
\multirow{4}{*}{ Retrieval}  &  \multirow{2}{=}{NL Code Search } & Comment-to-Snippet Search & 446K/22K/41K                   & Retrieve code snippet for given comment 
\\ \cmidrule{3-5}
                            &                                 & Problem-to-Program Search & 50K/3K/5K   & Retrieve program for given problem description                         
                            \\ \cmidrule{2-5}
                            &  \multirow{2}{=}{XL Code Search} & Snippet-to-Snippet Search & 872K/47K/83K                                  & Retrieve code snippets in other languages for given snippet   
                            \\ \cmidrule{3-5}
                            &                                 & Program-to-Program Search & 106K/6K/11K &  Retrieve programs in other languages for given snippet
                            \\ \bottomrule
\end{tabular}

\end{table*}

\section{Code Tasks}
The tasks can be divided into two categories: generation and retrieval. The generation tasks include Code Translation, Code Summarization, and Code Synthesis. The retrieval tasks include NL (natural language) Code Search and XL (Cross-Lingual) Code Search. All the tasks are at both snippet-level and program-level. Figure \ref{fig:data_task} shows the input and output data for each of the tasks. Table \ref{tab:task_overview} summarizes all the tasks introduced and some aggregate data statistics corresponding to each task.

\textbf{Code Translation (Code-to-Code):}
Code Translation is the problem of converting source code from one programming language to another. Efficient and accurate code translation is valuable in scenarios like legacy code migration, software platform adaptation, etc.
The proposed $\dname$ dataset provides parallel data in 7 common programming languages, supporting translation for 42 language pairs at both snippet and program level. 

\textbf{Code Summarization  (Code-to-Text):}
The objective of Code Summarization task is to generate natural language descriptions of the code that is given as input. We perform this task under two settings, generating snippet level summary by leveraging the comment-snippet pairings, and generating problem level summary using the problem description and program code pairings. Applications of this task include increasing the comprehensibility of uncommented or unfamiliar code to first time viewers and making it easier to collaborate as well as educate.

\textbf{Code Synthesis (Text-to-Code):}
The Code Synthesis task focuses on generating source code from text inputs. It includes Snippet Synthesis and Program Synthesis. We use the comment of each code snippet as input to generate the code snippet for the Snippet Synthesis task, since they are of similar length (as shown in Table \ref{tab:data_stats_small}). However, programs are usually much longer (Avg. 202 tokens) than problem descriptions (Avg. 11 tokens). To generate programs, it is necessary that the input text is detailed and informative. Therefore, we use a combination of problem description and step-by-step comments as input to generate the entire program. Since the programs in $\dname$ are well commented (9 comments/snippets per program on an average) 
this ensures that the models have enough information to synthesize the whole program.

\textbf{Code Search:}
The NL (Natural Language) Code Search in this paper refers to using text input to retrieve relevant code. The snippet and program level task use Comment and Problem Description as query, respectively. XL (Cross-lingual) Code Search is the task of retrieving code that performs similar functions in multiple other languages given a piece of code in one particular language. Unlike NL code search, using code as queries to search for similarly functioning code in a multilingual setting is relatively unexplored task. This task also includes both snippet and program level. To account for multiple correct answers, we use a modified \textit{MRR} (Mean Reciprocal Rank) for evaluation (details in Appendix A.6).


\section{Experiments}

All the baselines were initialized with the pretrained weights and default configuration (including hyper-parameters) released by the corresponding original authors of the works. We changed the source and target sequence lengths to align with the dataset based on the task. The models were trained using 4 RTX 8000 GPUs with 48GB memory on each GPU. The code for training and evaluation is released in the GitHub repository of the dataset.

\subsection{Evaluation Metrics and Baselines}
We use the following metrics to evaluate different tasks proposed in this work: (i) BLEU \citep{papineni2002bleu} score to evaluate code-to-text generation tasks;, (ii) BLEU and CodeBLEU\footnote{We extended the CodeBLEU metric to support C and C++. Related code is released in the GitHub repo.} \citep{ren2020codebleu} to evaluate code-to-code and text-to code generation tasks, and (iii) Mean Reciprocal Rank (MRR) to evaluate retrieval tasks.\\
We use the following models/methods for our comparison: \\
\textbf{Naive Copy} \cite{lu2021codexglue} directly copies the input source code as the output, which shows how similar two programming languages are. It is only used for translation tasks. \\
\textbf{RoBERTa} \citep{liu2019roberta} is a robustly optimized version of BERT pretrained on huge natural language corpora. We use it only for retrieval tasks.\\
\textbf{CodeBERT} \citep{feng2020codebert} uses the BERT \citep{devlin2019bert} architecture pretrained on CodeSearchNet \citep{husain2019codesearchnet} data.
We use the encoder-only version for retrieval tasks and encoder-decoder version (the decoder is randomly initialized) for generation tasks. \\
\textbf{PLBART} \citep{ahmad2021unified} is initialized with mBART \citep{liu2020multilingual} and further pretrained on a large-collection of Java and Python functions and natural language descriptions from Github and StackOverflow with denoising auto-encoding objective. 
\\
\textbf{CodeT5} \citep{wang2021codet5} employs T5 \citep{raffel2020exploring} architecture and is pretrained on corpora of 8 programming languages (Java, Python, C\#, JS, PHP, C, Ruby, Go) with identifier-aware objective. 

\subsection{Result Analysis}




Table \ref{tab:translation_results} 
shows the performance of baseline models for Code Translation, Code Synthesis, Code Summarization, and Code Search tasks.

\begin{table*}[]
 \aboverulesep=0ex
 \belowrulesep=0ex
\setlength{\tabcolsep}{4.6pt}
\renewcommand{\arraystretch}{1.4}
\scriptsize
\rmfamily

\caption{\label{tab:translation_results}From top to bottom, the table contains results for Code Translation, Code Synthesis, Code Summarization, and Code Search at the snippet-level and program-level. CodeBLEU scores are reported for Code Generation tasks (Translation and Synthesis). For Translation, the language column on the left represents the source language and the row on the top represents the target language. BLEU scores are reported for Summarization and MRR for Search.}

\begin{tabular}{ p{1.5cm}|P{1.2cm}|P{0.42cm}P{0.42cm}P{0.42cm}P{0.42cm}P{0.42cm}P{0.42cm}P{0.46cm}|P{0.42cm}P{0.42cm}P{0.42cm}P{0.42cm}P{0.42cm}P{0.42cm}P{0.46cm} }

\toprule\multicolumn{2}{ c|}{}             & \multicolumn{7}{c|}{\textbf{Snippet-level}}                                                                            & \multicolumn{7}{c }{\textbf{Program-level}}                                                                            \\
\midrule
\multicolumn{1}{c|}{\textbf{CodeBLEU}} & \textbf{Model} & \textbf{C++}   & \textbf{Java}  & \textbf{Py} & \textbf{C\#}   & \textbf{JS}    & \textbf{PHP}   & \textbf{C}     & \textbf{C++}   & \textbf{Java}  & \textbf{Py} & \textbf{C\#}   & \textbf{JS}    & \textbf{PHP}   & \textbf{C}     \\\midrule

\multirow{4}{*}{\rule{0pt}{1.5ex}\textbf{C++}}     & Naive Copy     &   \textendash & 64.56 & 34.79 & 63.19 & 53.16 & 42.56 & 84.2 & \textendash & 57.36 & 17.68 & 58.02 & 53.16 & 18.97 & 75.91  \\
                 & CodeBERT    &      \textendash & 84.94 & 74.55 & 84.99 & 82.79 & 68.56 & 45.46 & \textendash & 74.73 & 24.96 & 76.35 & 72.95 & 50.4 & 21.84 \\
                 & PLBART       &     \textendash & 83.85 & 74.89 & 84.57 & 83.19 & 68.62 & 83.95 & \textendash & 75.26 & 70.13 & 78.01 & 61.85 & 67.01 & 72.59        \\
                 & CodeT5           &      \textendash & \textbf{86.35} & \textbf{76.28} & \textbf{85.85} & \textbf{84.31} & \textbf{69.87} & \textbf{90.45} & \textendash & \textbf{80.03} & \textbf{71.56} & \textbf{81.73} & \textbf{79.48} & \textbf{70.44} & \textbf{85.67}          \\ \arrayrulecolor{black!50}\midrule

\multirow{4}{*}{\rule{0pt}{1.5ex}\textbf{Java}}     & Naive Copy     &   70.85 & \textendash & 35 & 78.43 & 57.81 & 42.49 & 69.74 & 64.25 & \textendash & 39.87 & 72.68 & 57.81 & 42.51 & 62.48  \\
                 & CodeBERT    &      87.27 & \textendash & 58.39 & 92.26 & 84.63 & 67.26 & 39.94 & 79.36 & \textendash & 8.51 & 84.43 & 76.02 & 51.42 & 21.22         \\
                 & PLBART       &     87.31 & \textendash & 58.3 & 90.78 & 85.42 & 67.44 & 72.47 & 81.41 & \textendash & 66.29 & 83.34 & 80.14 & 67.12 & 63.37        \\
                 & CodeT5   &    \textbf{88.26} & \textendash & \textbf{74.59} & \textbf{92.56} & \textbf{86.22} & \textbf{69.02} & \textbf{82.78} & \textbf{84.26} & \textendash & \textbf{69.57} & \textbf{87.79} & \textbf{80.67} & \textbf{69.44} & \textbf{78.78}         \\ \midrule
\multirow{4}{*}{\rule{0pt}{1.5ex}\textbf{Python}}      & Naive Copy     &   39.22 & 31.89 & \textendash & 31.79 & 38.34 & 36.02 & 37.79 & 37.47 & 29.78 & \textendash & 27.59 & 38.42 & 35.48 & 35.66  \\
                 & CodeBERT    &      80.46 & 58.5 & \textendash & 54.72 & 57.38 & 65.14 & 10.7 & 68.87 & 28.22 & \textendash & 17.8 & 23.65 & 49.3 & 18.32  \\
                 & PLBART       &     80.15 & 74.15 & \textendash & 73.5 & 73.2 & 66.12 & 62.15 & 74.38 & 67.8 & \textendash & 66.03 & 69.3 & 64.85 & 29.05 \\
                 & CodeT5           &      \textbf{81.56} & \textbf{78.61} & \textendash & \textbf{78.89} & \textbf{77.76} & \textbf{67.54} & \textbf{68.67} & \textbf{78.85} & \textbf{73.15} & \textendash & \textbf{73.35} & \textbf{71.8} & \textbf{67.5} & \textbf{56.35} \\ \midrule
\multirow{4}{*}{\rule{0pt}{1.5ex}\textbf{C\#}}     & Naive Copy     &   69.78 & 78.71 & 34.77 & \textendash & 57.85 & 42.53 & 66.73 & 64 & 73.63 & 40.09 & \textendash & 57.79 & 42.96 & 60.87  \\
                 & CodeBERT    & 86.96 & 90.15 & 56.92 & \textendash & 84.38 & 67.18 & 40.43 & 78.52 & 82.25 & 10.82 & \textendash & 75.46 & 51.76 & 21.63  \\
                 & PLBART       &     84.98 & 6.27 & 69.82 & \textendash & 85.02 & 67.3 & 75.74 & 80.17 & 81.37 & 67.02 & \textendash & 79.81 & 67.12 & 57.6   \\
                 & CodeT5           &      \textbf{88.06} & \textbf{91.69} & \textbf{73.85} & \textendash & \textbf{85.95} & \textbf{68.97} & \textbf{81.09} & \textbf{83.59} & \textbf{85.7} & \textbf{69.52} & \textendash & \textbf{80.5} & \textbf{69.63} & \textbf{77.35} \\ \midrule
\multirow{4}{*}{\rule{0pt}{1.5ex}\textbf{JS}}     & Naive Copy     &   60.82 & 59.25 & 38.84 & 64.27 & \textendash & 41.56 & 55.84 & 53.81 & 51.77 & 42.31 & 54.86 & \textendash & 42.11 & 49.04   \\
                 & CodeBERT    & 84.38 & 84.42 & 52.57 & 84.74 & \textendash & 66.66 & 33.29 & 75.43 & 72.33 & 9.19 & 75.47 & \textendash & 52.08 & 19.79  \\
                 & PLBART       &     84.45 & 84.9 & 69.29 & 85.05 & \textendash & 67.09 & 72.65 & 80.19 & 76.96 & 64.18 & 78.51 & \textendash & 67.24 & 67.7\\
                 & CodeT5           &      \textbf{85.06} & \textbf{85.48} & \textbf{73.15} & \textbf{85.96} & \textendash & \textbf{68.42} & \textbf{80.49} & \textbf{82.14} & \textbf{79.91} & \textbf{68.42} & \textbf{81.77} & \textendash & \textbf{68.76} & \textbf{74.57} \\ \midrule
\multirow{4}{*}{\rule{0pt}{1.5ex}\textbf{PHP}}     & Naive Copy     &   36.33 & 35.61 & 24.62 & 36.67 & 35.55 & \textendash & 35.95 & 34.62 & 31.33 & 25.68 & 32.81 & 32.26 & \textendash & 33.45  \\
                 & CodeBERT    & 82.58 & 81.57 & 69.29 & 80.96 & 79.94 & \textendash & 28.45 & 50.13 & 46.81 & 16.92 & 49.75 & 48.12 & \textendash & 22.19  \\
                 & PLBART       &     83.87 & 81.66 & 71.17 & 78 & 82.94 & \textendash & 57.39 & 79.4 & 72.77 & 61.26 & 74.16 & 44.26 & \textendash & 56.23  \\
                 & CodeT5   &      \textbf{86.33} & \textbf{85.12} & \textbf{73.22} & \textbf{84.56} & \textbf{83.56} & \textendash & \textbf{79.3} & \textbf{85.55} & \textbf{82.09} & \textbf{72.26} & \textbf{83.79} & \textbf{81.72} & \textendash & \textbf{65.86} \\ \midrule
\multirow{4}{*}{\rule{0pt}{1.5ex}\textbf{C}}     & Naive Copy     &   83.93 & 65.46 & 38.49 & 63.05 & 55.55 & 41.85 & \textendash & 78.4 & 59.41 & 20.2 & 59.83 & 53.54 & 19.75 & \textendash  \\
                 & CodeBERT    & 45.84 & 39.69 & 13.55 & 39.71 & 29.85 & 38.88 & \textendash & 21.7 & 21.27 & 21.1 & 19.5 & 15.64 & 31.71 & \textendash  \\
                 & PLBART       &     82.53 & 72.35 & 49.16 & 75.78 & 75.05 & 60.86 & \textendash & 78.42 & 13.45 & 5.53 & 45.15 & 31.47 & 25.17 & \textendash   \\
                 & CodeT5   &      \textbf{90.26} & \textbf{81.81} & \textbf{63.81} & \textbf{83.05} & \textbf{79.73} & \textbf{66.32} & \textendash & \textbf{88.17} & \textbf{76.12} & \textbf{56.32} & \textbf{80.2} & \textbf{76.5} & \textbf{64.28} & \textendash \\  \arrayrulecolor{black}\hline
     \multicolumn{1}{c|}{}             & \multicolumn{1}{c|}{}                                                                            & \multicolumn{7}{c|}{} & \multicolumn{7}{c}{}\\[-2.7ex]
     \hline
\multicolumn{1}{c|}{\textbf{CodeBLEU}} &  \textbf{Model} & \textbf{C++}   & \textbf{Java}  & \textbf{Py} & \textbf{C\#}   & \textbf{JS}    & \textbf{PHP}   & \textbf{C}     & \textbf{C++}   & \textbf{Java}  & \textbf{Py} & \textbf{C\#}   & \textbf{JS}    & \textbf{PHP}   & \textbf{C}   \\ \midrule
\multirow{3}{1.2cm}{\rule{0pt}{1.5ex}Code Synthesis}    & CodeBERT        & 22.7 & 25.53 & 12.26 & 23.44 & 23.87 & 36.47 & 10.63 & 26.51 & 31.14 & 24.5 & 33.37 & 29.09 & 39.84 & 18.08\\                       &     PLBART      & 34.89 & 32.23 & 4.62 & 29.36 & 29.63 & \textbf{37.56} & \textbf{22.88} & 44.09 & 41.55 & 33.77 & 40.7 & 38.33 & 43.01 & 6.72  \\                       & CodeT5           & \textbf{35.48} & \textbf{33.51} & \textbf{21.1} & \textbf{30.64} & \textbf{29.99} & 36.37 & 21.93 & \textbf{45.18} & \textbf{42.73} & \textbf{35.02} & \textbf{43.6} & \textbf{38.66} & \textbf{45.02} & \textbf{34.88} \\ \hline
     \multicolumn{1}{c|}{}             & \multicolumn{1}{c|}{}                                                                            & \multicolumn{7}{c|}{} & \multicolumn{7}{c}{}\\[-2.7ex]
     \hline   
\multicolumn{1}{c|}{\textbf{BLEU}} &  \textbf{Model} & \textbf{C++}   & \textbf{Java}  & \textbf{Py} & \textbf{C\#}   & \textbf{JS}    & \textbf{PHP}   & \textbf{C}     & \textbf{C++}   & \textbf{Java}  & \textbf{Py} & \textbf{C\#}   & \textbf{JS}    & \textbf{PHP}   & \textbf{C}\\ \midrule
\multirow{3}{1.5cm}{\rule{0pt}{1.5ex}Code Summarization}  & CodeBERT        & 14.4 & 13.13 & 3.96 & 14.07 & 11.81 & 11.25 & 5.84 & 7.68 & 5.47 & 2.04 & 7.58 & 7.67 & 7.5 & 6.64 \\                       & PLBART  & 14.77 & 13.76 & 8 & 14.37 & 10.93 & 9.07 & \textbf{7.5} & 7.65 & 6.35 & 4.86 & \textbf{9.23} & 6.78 & 6.03 & 4.14  \\                       &    CodeT5       & \textbf{17.36} & \textbf{16.69} & \textbf{10.76} & \textbf{17.44} & \textbf{14.34} & \textbf{13.42} & 6.63 & \textbf{9.62} & \textbf{8.82} & \textbf{6.32} & 7.75 & \textbf{8.23} & \textbf{10.5} & \textbf{12.84}  \\ 
               \hline
     \multicolumn{1}{c|}{}             & \multicolumn{1}{c|}{}                                                                            & \multicolumn{7}{c|}{} & \multicolumn{7}{c}{}\\[-2.7ex]
     \hline      

\multicolumn{1}{c|}{\textbf{MRR}} &  \textbf{Model} & \textbf{C++}   & \textbf{Java}  & \textbf{Py} & \textbf{C\#}   & \textbf{JS}    & \textbf{PHP}   & \textbf{C}     & \textbf{C++}   & \textbf{Java}  & \textbf{Py} & \textbf{C\#}   & \textbf{JS}    & \textbf{PHP}   & \textbf{C}   \\ \midrule
\multirow{2}{1.5cm}{NL Code Search}     & RoBERTa         &25.77 & 25.85 & 27.08 & 25.64 & 26.78 & 33.47 & 36.14 & 51.47 & 50.4 & 48.98 & 52.24 & 50.05 & 62.01 & \textbf{56.34}\\                      & CodeBERT        & \textbf{29.77} & \textbf{29.41} & \textbf{30.94} & \textbf{29.08} & \textbf{31.2} & \textbf{38.75} & \textbf{41.56} & \textbf{59.13} & \textbf{56.07} &\textbf{57.97} & \textbf{56.65} & \textbf{54.37} & \textbf{65.13} & 47.13\\ \arrayrulecolor{black!30}\midrule
 \multirow{2}{1.5cm}{XL Code Search}     & RoBERTa         & 41.73 & 41.25 & 36.16 & 41.18 & 43.17 & \textbf{41.17} & 37.1 & 48.28 & 47.66 & 46.11 & 46.4 & 47.6 & 43.76 & 40.15\\                    & CodeBERT        & \textbf{42.11} & \textbf{41.71} & \textbf{36.98} & \textbf{41.52} & \textbf{43.41} & 41.09 & \textbf{37.87} & \textbf{48.71} & \textbf{48.33} & \textbf{47.24} & \textbf{47.96} & \textbf{47.66} & \textbf{44.02} & \textbf{40.43} \\
 \arrayrulecolor{black}\bottomrule
\end{tabular}

\end{table*}

\begin{table*}[]
 \aboverulesep=0ex
 \belowrulesep=0ex
\setlength{\tabcolsep}{3.7pt}
\renewcommand{\arraystretch}{1.15}
\scriptsize
\rmfamily

\caption{\label{tab:translation_transfer} Transfer learning from Snippet-Level training for Program Translation task on low resource language C. ST - Snippet Transfer.}

\begin{tabular}{ p{1.7cm}|P{0.7cm}P{0.7cm}P{0.7cm}P{0.7cm}P{0.7cm}P{0.7cm}P{0.8cm}P{0.7cm}P{0.7cm}P{0.7cm}P{0.7cm}P{0.7cm}P{0.7cm} }

\toprule
\multicolumn{1}{c|}{\textbf{Model}} & \multicolumn{1}{c}{\textbf{C-C++}} & \multicolumn{1}{c}{\textbf{C-Java}} & \multicolumn{1}{c}{\textbf{C-Py}} & \multicolumn{1}{c}{\textbf{C-C\#}} & \multicolumn{1}{c}{\textbf{C-JS}} & \multicolumn{1}{c}{\textbf{C-PHP}} & \multicolumn{1}{c}{\textbf{C++-C}} & \multicolumn{1}{c}{\textbf{Java-C}} & \multicolumn{1}{c}{\textbf{Py-C}} & \multicolumn{1}{c}{\textbf{C\#-C}} & \multicolumn{1}{c}{\textbf{JS-C}} & \multicolumn{1}{c}{\textbf{PHP-C}} \\ \midrule
CodeBERT & 21.67 & 21.27 & 21.1 & 19.48 & 15.68 & 31.71 & 21.87 & 21.27 & 18.32 & 21.57 & 19.79 & 22.19 \\
CodeBERT+ST & \textbf{38.85} & \textbf{37.55} & \textbf{19.79} & \textbf{33.52} & \textbf{27.1} & \textbf{37.61} & \textbf{31.99} & \textbf{30.52} & \textbf{24.07} & \textbf{34.16} & \textbf{29.67} & \textbf{28.35} \\
PLBART & 78.42 & 13.43 & 5.53 & 45.14 & 31.42 & 25.17 & 72.61 & 63.4 & 29.01 & 57.6 & 67.71 & 56.15 \\
PLBART+ST & \textbf{81.1} & \textbf{70.78} & \textbf{44.26} & \textbf{72.68} & \textbf{73.27} & \textbf{60.71} & \textbf{79.72} & \textbf{77.3} & \textbf{47.48} & \textbf{74.09} & \textbf{72.6} & \textbf{64.64} \\
CodeT5 & 88.17 & 76.15 & 56.3 & 80.2 & 76.42 & 64.28 & 85.67 & 78.76 & 56.44 & 77.38 & 74.56 & 65.8 \\
CodeT5+ST & \textbf{89.06} & \textbf{79.04} & \textbf{62.61} & \textbf{80.53} & \textbf{78.59} & \textbf{68.31} & \textbf{88.96} & \textbf{82.08} & \textbf{60.97} & \textbf{80.93} & \textbf{79.58} & \textbf{77.58}\\
 \arrayrulecolor{black}\bottomrule
\end{tabular}

\end{table*}

\begin{table*}[]
\centering
\tiny
\rmfamily
 \aboverulesep=0ex
 \belowrulesep=0ex
  \renewcommand{\arraystretch}{1.3}
\setlength{\tabcolsep}{1.4pt}

\caption{\label{tab:compilation_errors} Top compilation errors in each target language (Javascript not included).}

\begin{tabular}{p{1cm}|p{4.3cm}|p{4.2cm}|p{4cm}}
\toprule

\multicolumn{1}{c|}{\textbf{\scriptsize Language}} & \multicolumn{3}{c}{\textbf{\scriptsize Top-3 Compilation Errors in Each Target Language}}     
\\ \midrule

\textbf{C++}	& expected ‘\}’ at end of input &	stray ‘\#’ in program &	‘define’ does not name a type\\ \midrule
\textbf{Java} &	';' expected &	not a statement &	unclosed character literal \\ \midrule
\textbf{Python} &	SyntaxError: invalid syntax &	SyntaxError: unexpected EOF while parsing &	IndentationError: expected an indented block \\ \midrule
\textbf{C\#} &	Too many characters in character literal &	Unexpected symbol `end-of-file' &	Newline in constant \\ \midrule
\textbf{PHP} &	Syntax error, unexpected '\}',expecting EOF.. &	Syntax error, unexpected ')'.. &	Syntax error, unexpected EOF on line 1 \\ \midrule
\textbf{C} &	expected declaration or statement at end of input &	expected ‘=’,‘,’,‘;’,‘asm’ ... before ‘)’ token &	expected statement before ‘)’ token \\
 \bottomrule
\end{tabular}

\end{table*}


\textbf{Effect of Sequence-to-Sequence Pretraining:}
In Table \ref{tab:translation_results}, on an average, CodeBERT performs significantly worse than PLBART and CodeT5 on almost all the generation tasks (refer to the first three sections of the table). Different from PLBART and CodeT5, which are both encoder-decoder models pretrained with sequence-to-sequence objectives, only the encoder in CodeBERT is pretrained, and the decoder weights are randomly initialized for sequence-to-sequence tasks. Experimental results show that encoder-decoder architecture and sequence-to-sequence pretraining are better aligned with generation tasks and thus can potentially achieve superior performance.

\textbf{Effect of Pretraining on Specific Languages:}
CodeBERT is pretrained on CodeSearchNet, which contains data from 6 programming languages, Java, Python, Javascript, PHP, Ruby, and Go. PLBART is pretrained on Java and Python from GitHub data. CodeT5 is trained on the 6 languages from CodeSearchNet and additional C and C\#. In Table \ref{tab:translation_results}, CodeT5 consistently outperforms the other two models for almost all generation tasks. When the source or target language is C, CodeT5 outperforms the other two by a wide margin. Pre-training on specific languages can potentially benefit the generation tasks with these languages as either input or output.

\textbf{Performance on Low-Resource Languages:}
In Table \ref{tab:translation_results}, most models performs significantly worse on C compared to other languages, both when C is source or target language, in almost all the tasks (except for Code Search). As shown in Table \ref{tab:data_stats_small}, C has the least number of samples for all the tasks. It shows that tasks in low-resource languages are potentially more challenging. 

\textbf{Effect of Transfer Learning from Snippet-level Training:} 
From Table \ref{tab:translation_results}, first section, we noticed that models perform significantly better at snippet-level than program-level on most language pairs in the translation task. 
This is because 1) Snippets are much shorter than programs. As shown in Table \ref{tab:data_stats_small}, the average length of snippets is 1/7 of the programs.  2) Snippet data is much more than program data. As shown in Table \ref{tab:task_overview}, the amount of pairwise snippet data is 8 times of program data. 
Motivated by this, 
we employ transfer learning from snippet-level training to improve the Program Translation performance on low-resource language C. Table \ref{tab:translation_transfer} shows the performance of each model with and without the transfer learning. For example, CodeBERT is trained only on program data; "CodeBERT + ST" (ST is short for Snippet Transfer) model is first trained on the snippet data, and then on program data. All the models' performances improve by a wide margin  on all the language pairs after snippet-level transfer learning, both when C is the source or target language.

\textbf{Top Compilation Errors in Generated Programs:} Table \ref{tab:compilation_errors} shows the top compilation error types from compiling the generated programs from the Program Translation task. We aggregated the results of generated programs from all the baselines by the target language, because 1) the top error types of each baseline are very similar and 2) the space is limited. From this table, we can see that the top error types are mostly syntactic errors, such as bracket mismatch (C++, PHP, C), indentation mismatch (Python), missing ';' (Java). This indicates that the models need improvement in capturing the structure of the programs. 

\subsection{Limitations and Future Work}
From our analysis of the results, we can conclude that Sequence-to-Sequence pretraining tasks, multilingual pretraining data, and Snippet-level Transfer Learning can potentially improve the performance on multiple tasks and low resource languages. This is an important insight for the design and development of future models in this domain. A good code generation model should also be able to learn and preserve the structure of the code since the current models mostly make syntactic errors in generation. For the evaluation of code generation tasks, we use CodeBLEU as metric, which evaluates the code syntax and semantics along with $n$-gram matching (as in BLEU). However, the evaluation can be further improved by using test cases. Automated test case generation can be explored in future work. The tasks we introduce aim to rigorously evaluate code models with the parallel data from the dataset. Therefore, not all the tasks have practical applications in real-world, especially the snippet-level tasks. One future direction is to make use of the comments and snippets to iteratively generate programs. 

\section{Conclusion}

In this paper, we introduce a new dataset which is parallel across 8 languages (7 programming languages and 1 natural language) at both snippet level and program level. To the best of our knoweldge, it is the largest parallel dataset for source code in terms of both size and number of languages. We also introduce 10 different cross-lingual tasks to facilitate the development and evaluation of models in this domain. Moreover, we run experiments for all the 10 tasks on the proposed dataset with a number of state-of-the-art baseline models and provided insights about model design for the new challenges. We believe that this dataset will be of significant value to the research community and can potentially benefit a number of code-related research problems.

\bibliography{neurips_data_2022}
\bibliographystyle{acl_natbib}

\appendix
\newpage
\section{APPENDIX}

\subsection{BLEU Scores for Code Generation}
Due to space constraints in the main document, we are including the BLEU results for Code Translation and Synthesis tasks in Table \ref{tab:translation_results_codebleu} in the appendix. 

BLEU score has been the defacto evaluation metric used to evaluate natural language translation tasks. It measures the similarity between the generated translation and a set of reference texts. However, different from natural languages, programming languages have more rigorous syntax and semantics. A minor change in the code sequence, such as addition or removal of a bracket, may not affect the BLEU score by much, but it can potentially alter the structure and functionality of the code substantially. Therefore, in the main paper, we use CodeBLEU as the evaluation metric for code generation task, as it takes into consideration the Abstract Syntax Tree (AST) matching and Data Flow Graph (DFG) matching, which measure the syntax and semantics of the code, respectively.


Due to the fact that BLEU only measures the $n$-gram matching and ignores code syntax and semantics, it can be observed from our translation results (see Table \ref{tab:translation_results_codebleu}) that BLEU scores clearly over estimate the model performance. Almost all BLEU score values can be found to be greater than CodeBLEU scores presented in the main paper. 
The effect is more pronounced for the program level tasks than for the snippet level tasks. This shows that maintaining long term structure in full programs, which are much longer than snippets, is harder.

However, from our Code Synthesis results (see Table \ref{tab:translation_results_codebleu}), we observe that the BLEU scores are lower than CodeBLEU scores. This is because the Code Synthesis results are much lower (in absolute value) than translation results to begin with, and taking into account AST and DFG matching increases the overall scores.

\begin{table*}[hb]

\caption{\label{tab:translation_results_codebleu}BLEU scores for two tasks (Translation and Code Synthesis) separated by a double horizontal rule. Above the double rule are presented BLEU scores of the Translation task for the 42 programming language pairs in the $\dname$ dataset. Below the double rule are the BLEU scores for the Code Synthesis task. Column headers represent target languages. For translation, row headers represent source languages.}

 \aboverulesep=0ex
 \belowrulesep=0ex
\setlength{\tabcolsep}{4.6pt}
\renewcommand{\arraystretch}{1.3}
\scriptsize
\rmfamily
\begin{tabular}{ p{1.1cm}|P{1.2cm}|P{0.45cm}P{0.45cm}P{0.45cm}P{0.45cm}P{0.45cm}P{0.45cm}P{0.46cm}|P{0.45cm}P{0.45cm}P{0.45cm}P{0.45cm}P{0.45cm}P{0.45cm}P{0.46cm} }

\toprule\multicolumn{2}{ c|}{}             & \multicolumn{7}{c|}{\textbf{Snippet-level}}                                                                            & \multicolumn{7}{c }{\textbf{Program-level}}                                                                            \\
\midrule
\multicolumn{1}{c|}{\textbf{BLEU}}  & \textbf{Model} & \textbf{C++}   & \textbf{Java}  & \textbf{Py} & \textbf{C\#}   & \textbf{JS}    & \textbf{PHP}   & \textbf{C}     & \textbf{C++}   & \textbf{Java}  & \textbf{Py} & \textbf{C\#}   & \textbf{JS}    & \textbf{PHP}   & \textbf{C}     \\\midrule

\multirow{4}{*}{\rule{0pt}{1.5ex}\textbf{C++}}     & Naive Copy     &   \textendash & 64.57 & 37.29 & 65.89 & 59.73 & 37.44 & 84.44 & \textendash & 64.47 & 34.48 & 65.98 & 58.09 & 38.13 & 84  \\
                 & CodeBERT    &      \textendash & 85.03 & 79.72 & 85.64 & 84.61 & 87.18 & 44.48 & \textendash & 80.09 & 15.43 & 81.24 & 78.14 & 50.68 & 11.7 \\
                 & PLBART       & \textendash & 84.02 & 80.12 & 84.86 & 85.34 & 87.31 & 85.26 & \textendash & 81.23 & 77.5 & 83.96 & 69.6 & 83.94 & 77.94       \\
                 & CodeT5   & \textendash & \textbf{86.45} & \textbf{81.73} & \textbf{86.55} & \textbf{86.24} & \textbf{89.84} & \textbf{91.82} & \textendash & \textbf{84.29} & \textbf{78.69} & \textbf{85.69} & \textbf{83.75} & \textbf{90.88} & \textbf{93.64}      \\ \arrayrulecolor{black!50}\midrule

\multirow{4}{*}{\rule{0pt}{1.5ex}\textbf{Java}}     & Naive Copy     &   64.48 & \textendash & 34.04 & 76.95 & 60.42 & 35.02 & 65.67 & 64.88 & \textendash & 31.28 & 78.09 & 58.23 & 34.35 & 65.48  \\
                 & CodeBERT    &      88.18 & \textendash & 61.45 & 92.64 & 86.42 & 84.57 & 38.47 & 83.24 & \textendash & 2.51 & 88.31 & 81.16 & 52.7 & 12.27        \\
                 & PLBART       &  87.69 & \textendash & 55.05 & 91.63 & 87.17 & 84.92 & 71.43 & 85.83 & \textendash & 73.16 & 89 & 84.62 & 84.18 & 66.1    \\
                 & CodeT5   & \textbf{89.2} & \textendash & \textbf{79.37} & \textbf{92.9} & \textbf{87.95} & \textbf{88.12} & \textbf{84.29} & \textbf{87.67} & \textendash & \textbf{76.92} & \textbf{90.5} & \textbf{85.38} & \textbf{88.86} & \textbf{85.64}  \\ \midrule

\multirow{4}{*}{\rule{0pt}{1.5ex}\textbf{Python}}      & Naive Copy     &   37.36 & 34 & \textendash & 35.06 & 42.64 & 22.05 & 38.5 & 34.43 & 30.81 & \textendash & 31.93 & 39.57 & 21.22 & 35.68 \\
                 & CodeBERT    &  82.02 & 57.17 & \textendash & 53.94 & 57.64 & 80.32 & 9.01 & 72.56 & 25.24 & \textendash & 8.39 & 17.97 & 48.46 & 4.5 \\
                 & PLBART       &   81.17 & 69.17 & \textendash & 69.29 & 68 & 82.27 & 57.8 & 78.4 & 73.1 & \textendash & 71.14 & 73.73 & 79.56 & 27.07 \\
                 & CodeT5  & \textbf{83.07} & \textbf{76.61} & \textendash & \textbf{78.59} & \textbf{78.16} & \textbf{85.14} & \textbf{70.64} & \textbf{81.8} & \textbf{77.9} & \textendash & \textbf{78.21} & \textbf{76.25} & \textbf{85} & \textbf{51.89}\\ \midrule

\multirow{4}{*}{\rule{0pt}{1.5ex}\textbf{C\#}}     & Naive Copy     &  65.71 & 77.11 & 34.99 & \textendash & 61.13 & 35.25 & 66.63 & 66.13 & 78.04 & 32.2 & \textendash & 59.09 & 36.08 & 66.39 \\
                 & CodeBERT    & 87.76 & 89.99 & 59.63 & \textendash & 86.13 & 84.41 & 39.19 & 82.48 & 86.86 & 4.68 & \textendash & 80.8 & 53.39 & 11.48 \\
                 & PLBART       &   87.03 & 4.97 & 70.85 & \textendash & 86.88 & 84.66 & 76.83 & 84.77 & 87 & 73.94 & \textendash & 84.55 & 84.2 & 60.27  \\ 
                 & CodeT5 &\textbf{88.81} & \textbf{91.67} & \textbf{77.64} & \textendash & \textbf{87.75} & \textbf{88.01} & \textbf{83.54} & \textbf{87.18} & \textbf{89.28} & \textbf{76.81} & \textendash & \textbf{85} & \textbf{89.25} & \textbf{83.72}\\ \midrule

\multirow{4}{*}{\rule{0pt}{1.5ex}\textbf{JS}}     & Naive Copy     &  59.84 & 60.36 & 42.63 & 61.23 & \textendash & 33.05 & 56.07 & 57.99 & 57.26 & 39.58 & 58.54 & \textendash & 34 & 53.8 \\
                 & CodeBERT    & 84.97 & 84.1 & 54.19 & 84.69 & \textendash & 83.37 & 31.68 & 79.54 & 77.97 & 3.36 & 80.73 & \textendash & 54.02 & 11.04  \\
                 & PLBART       &    84.75 & 84.44 & 70.99 & 84.85 & \textendash & 84.19 & 73.57 & 84.24 & 82.46 & 70.68 & 83.86 & \textendash & 84.4 & 69.46\\
                 & CodeT5           &  \textbf{85.74} & \textbf{85.22} & \textbf{77.46} & \textbf{85.96} & \textendash & \textbf{86.91} & \textbf{82.03} & \textbf{85.72} & \textbf{84.14} & \textbf{75.46} & \textbf{85.44} & \textendash & \textbf{87.5} & \textbf{77.92}
 \\ \midrule

\multirow{4}{*}{\rule{0pt}{1.5ex}\textbf{PHP}}     & Naive Copy     & 37.39 & 35.03 & 22.08 & 35.2 & 33.17 & \textendash & 36.62 & 38.15 & 34.32 & 21.36 & 36.08 & 34.44 & \textendash & 37.53  \\
                 & CodeBERT    & 84.56 & 82.35 & 74.68 & 81.93 & 82.82 & \textendash & 27.94 & 52.19 & 50.94 & 10.87 & 54.61 & 53.13 & \textendash & 8.42  \\
                 & PLBART       &  84.81 & 82.13 & 76.42 & 78.86 & 85.43 & \textendash & 53.62 & 84.56 & 78.61 & 68.9 & 80.26 & 44.71 & \textendash & 56.16  \\
                 & CodeT5   &  \textbf{88.59} & \textbf{85.88} & \textbf{79.47} & \textbf{85.67} & \textbf{86.46} & \textendash & \textbf{83.05} & \textbf{89.96} & \textbf{86.95} & \textbf{80.29} & \textbf{87.77} & \textbf{87.01} & \textendash & \textbf{68.74}
\\ \midrule

\multirow{4}{*}{\rule{0pt}{1.5ex}\textbf{C}}     & Naive Copy     & 84.34 & 65.65 & 38.47 & 66.64 & 56.19 & 36.67 & \textendash & 83.99 & 65.29 & 35.94 & 66.4 & 54.52 & 37.53 & \textendash  \\
                 & CodeBERT    & 45.55 & 38.79 & 9.83 & 39.09 & 26.85 & 27.03 & \textendash & 15.51 & 17.77 & 6.01 & 14.92 & 13.06 & 8.53 & \textendash \\
                 & PLBART       &   83.01 & 72.21 & 44.76 & 76.26 & 78.8 & 72.37 & \textendash & 84.88 & 10.65 & 4.02 & 38.53 & 18.6 & 0.2 & \textendash  \\
     & CodeT5   & \textbf{91.76} & \textbf{82.12} & \textbf{65.89} & \textbf{84.06} & \textbf{82.16} & \textbf{82.82} & \textendash & \textbf{93.15} & \textbf{83.08} & \textbf{54.6} & \textbf{85.39} & \textbf{82.42} & \textbf{78.7} & \textendash      \\\arrayrulecolor{black}\hline
     \multicolumn{1}{c|}{}             & \multicolumn{1}{c|}{}                                                                            & \multicolumn{7}{c|}{} & \multicolumn{7}{c}{}\\[-2.7ex]
     \hline

\multicolumn{1}{c|}{\textbf{BLEU}} &  \textbf{Model} & \textbf{C++}   & \textbf{Java}  & \textbf{Py} & \textbf{C\#}   & \textbf{JS}    & \textbf{PHP}   & \textbf{C}     & \textbf{C++}   & \textbf{Java}  & \textbf{Py} & \textbf{C\#}   & \textbf{JS}    & \textbf{PHP}   & \textbf{C}   \\ \midrule
\multirow{3}{1.2cm}{\rule{0pt}{1.5ex}\textbf{Code Synthesis}}    & CodeBERT        & 17.19 & 18.78 & 7.82 & 18.58 & 19.53 & \textbf{22.54} & 5.16 & 21.39 & 27.16 & 19.58 & 30.83 & 25.53 & 29.6 & 8.85 \\                      &     PLBART      & 24.01 & 28.12 & 1.31 & 26.61 & 17.27 & 20.16 & \textbf{12.9} & 39.94 & 41.01 & 29.92 & 38.92 & 35.95 & 35.91 & 3.53 \\  & CodeT5           & \textbf{28.52} & \textbf{29.65} & \textbf{12.87} & \textbf{28.16} & \textbf{21.54} & 18.7 & 12.47 & \textbf{41.53} & \textbf{42.37} & \textbf{31.9} & \textbf{42.14} & \textbf{36.31} & \textbf{39.97} & \textbf{28.64} \\
\arrayrulecolor{black}\bottomrule
\end{tabular}
\end{table*}

\subsection{Dataset Statistics}
A detailed breakdown of the data statistics for the translation task is provided in this section due to space limitation in the main paper. Table \ref{tab:snip_prog_stats} summarizes the number of aligned code pairs contained in the train, validation, and test sets for all possible language pair combinations both at the snippet and program level. 

\begin{table}[]
 \aboverulesep=0ex
 \belowrulesep=0ex
 \renewcommand{\arraystretch}{1.1}
\setlength{\tabcolsep}{3.2pt}
\scriptsize
\rmfamily
\centering

\caption{\label{tab:snip_prog_stats}Number of pairwise code-code data in training, validation, and testing splits for each language-pair. The upper triangle (in bold font) shows the number of parallel code snippets, and the lower triangle shows the number of parallel programs. This data is used for the Code Translation and XL Code Search tasks. (\textbf{Py} is short for Python. \textbf{JS} is short for Javascript.)}

\begin{tabular}{ P{1cm}|P{0.8cm}|P{0.7cm}P{0.7cm}P{0.7cm}P{0.7cm}P{0.7cm}P{0.7cm}P{0.7cm}}

\toprule
\textbf{Lang} & \textbf{} & \textbf{C++}   & \textbf{Java}  & \textbf{Py} & \textbf{C\#}   & \textbf{JS}    & \textbf{PHP}   & \textbf{C} \\ \midrule
\multirow{3}{*}{\rule{0pt}{1.5ex}\textbf{C++}}     & train & \textbf{\textendash} & \textbf{89040} & \textbf{80100} & \textbf{85662} & \textbf{69507} & \textbf{17811} & \textbf{3386} \\
 & val & \textbf{\textendash} & \textbf{4419} & \textbf{3913} & \textbf{4408} & \textbf{3808} & \textbf{923} & \textbf{352} \\
 & test & \textbf{\textendash} & \textbf{8059} & \textbf{7228} & \textbf{7922} & \textbf{6965} & \textbf{1647} & \textbf{222} \\ \arrayrulecolor{black!30}\midrule
\multirow{3}{*}{\rule{0pt}{1.5ex}\textbf{Java}}     & train & 9450 & \textendash & \textbf{77759} & \textbf{87065} & \textbf{69341} & \textbf{17853} & \textbf{2996} \\
 & val & 490 & \textendash & \textbf{3938} & \textbf{4437} & \textbf{3826} & \textbf{929} & \textbf{353} \\
 & test & 901 & \textendash & \textbf{7259} & \textbf{8011} & \textbf{7005} & \textbf{1672} & \textbf{238} \\\midrule
 \multirow{3}{*}{\rule{0pt}{1.5ex}\textbf{Py}}       & train & 9139 & 8991 & \textendash & \textbf{75843} & \textbf{67219} & \textbf{17616} & \textbf{2478} \\
 & val & 468 & 471 & \textendash & \textbf{3922} & \textbf{3750} & \textbf{923} & \textbf{311} \\
 & test & 878 & 882 & \textendash & \textbf{7215} & \textbf{6861} & \textbf{1655} & \textbf{203} \\\midrule
\multirow{3}{*}{\rule{0pt}{1.5ex}\textbf{C\#}}      & train & 9187 & 9301 & 8826 & \textendash & \textbf{68093} & \textbf{17873} & \textbf{2958} \\
 & val & 488 & 491 & 470 & \textendash & \textbf{3826} & \textbf{928} & \textbf{352} \\
 & test & 890 & 898 & 877 & \textendash & \textbf{6961} & \textbf{1668} & \textbf{238} \\\midrule
 \multirow{3}{*}{\rule{0pt}{1.5ex}\textbf{JS}}     & train & 8482 & 8470 & 8182 & 8367 & \textendash & \textbf{17117} & \textbf{1875} \\
 & val & 472 & 475 & 459 & 475 & \textendash & \textbf{921} & \textbf{309} \\
 & test & 878 & 881 & 864 & 877 & \textendash & \textbf{1617} & \textbf{200} \\\midrule
 \multirow{3}{*}{\rule{0pt}{1.5ex}\textbf{PHP}}      & train & 3056 & 3068 & 3003 & 3071 & 2971 & \textendash & \textbf{856} \\
 & val & 157 & 158 & 153 & 158 & 157 & \textendash & \textbf{271} \\
 & test & 303 & 307 & 304 & 307 & 302 & \textendash & \textbf{183} \\\midrule
  \multirow{3}{*}{\rule{0pt}{1.5ex}\textbf{C}}      & train & 402 & 409 & 380 & 394 & 308 & 170 & \textendash \\
 & val & 59 & 59 & 59 & 59 & 59 & 55 & \textendash \\
 & test & 45 & 49 & 48 & 49 & 49 & 43 & \textendash \\
\arrayrulecolor{black}\bottomrule
\end{tabular}

\end{table}

\subsection{More Details about Data Collection}

The data was scraped from different sub-pages of the GeeksForGeeks website. A majority of the problems on this site belong to the following two categories: Data Structures and Algorithms. These two categories have different sub-categories within them. For example, the Data Structures page has the hyperlinked sub-categories of Array, Linked Lists, Stack, Queue, etc. which when clicked direct the user to all the problems relating to that specific sub-category and their corresponding solutions in different programming languages. The same goes for the Algorithms page. 

\begin{figure}[h]
\centering
\includegraphics[width=0.7\textwidth]{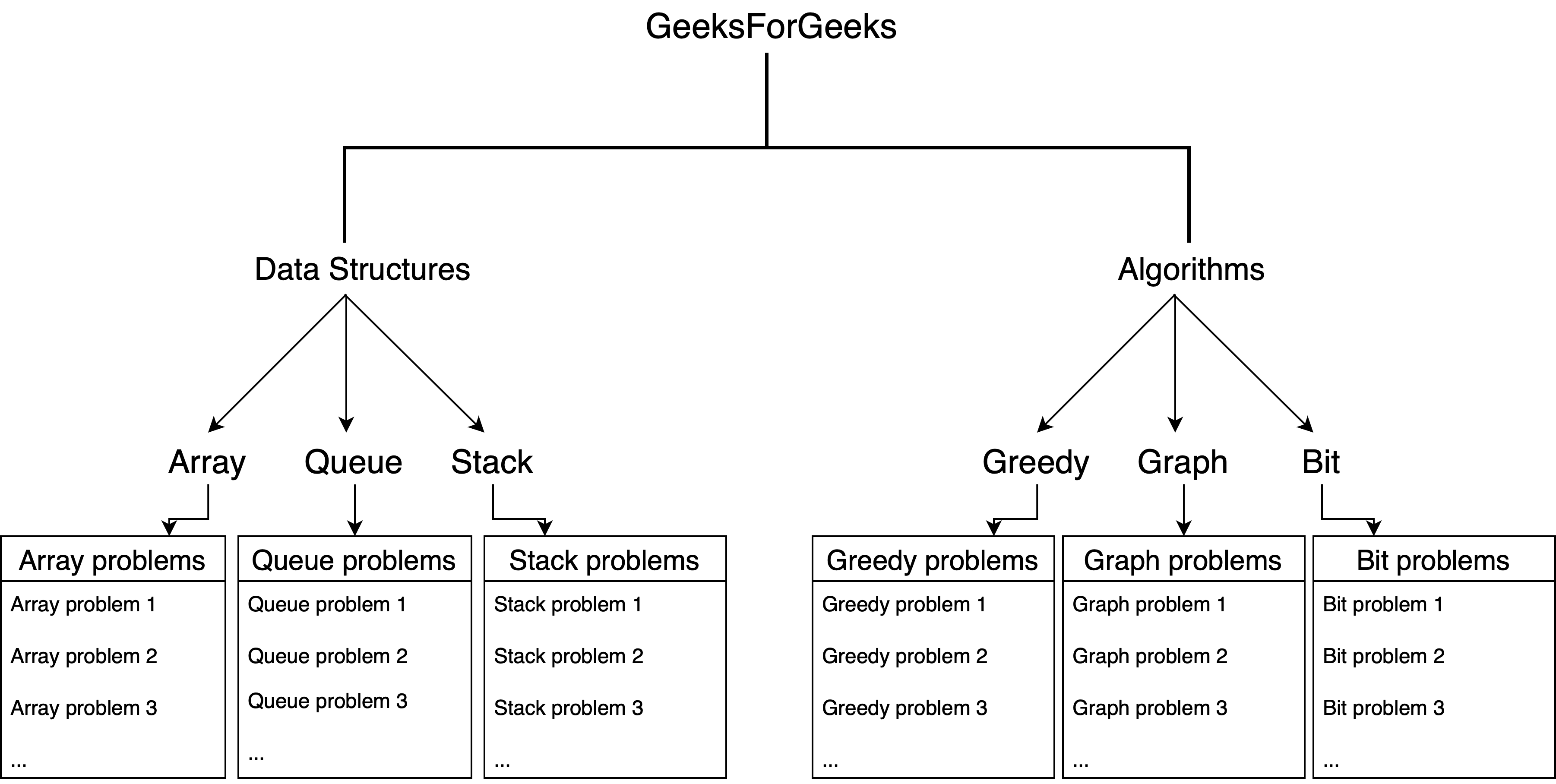}
\caption{A breakdown of the problem sets included in the data and their organization on the GeeksForGeeks portal. Only three sub-categories per category is shown here for purposes of brevity.}
\label{fig:g4gflow}
\end{figure}


For scraping the data, we used python scripts and some external libraries, the most important of them being BeautifulSoup4\footnote{https://www.crummy.com/software/BeautifulSoup/} and Selenium\footnote{https://www.selenium.dev/}. BeautifulSoup4 is a python library that facilitates the acquisition of data from HTML and XML files. Selenium is a python package that, in our case, is used to automate web browser functions to navigate through the GeeksForGeeks page directories. Each page in GFG that houses a problem and its solution has a uniform HTML page structure. This structure allows us to extract specifically targeted sections which are needed for our dataset using BeautifulSoup4. 

Using the directory structure of the GeekForGeeks website, the ability of Selenium to navigate through these pages and the utility of BeautifulSoup4 for content extraction from these pages, we could extract the data from the website with practically no manual intervention. Every problem page can have one or more solutions to the same problem in different languages. For example, if there is a problem statement “Given a number n, print n-th Fibonacci Number”, there can be different logics to solve the same problem. One solution may use recursion-based logic, another can utilize dynamic programming and yet another can use a space complexity optimized method to solve the same problem. Each logic has code in different languages and code pertaining to each logic resides in separate sections of the page which can be identified via their HTML tags. We extracted every possible problem statement and solution from the above-mentioned two categories and did not put any filter on what category, type, difficulty, etc. the solutions belong to at collection time.

\subsection{Creating Data Splits}
A natural way to generate train-validation-test sets from the data is to split at the problem level. However, the number of programs in different languages are imbalanced. Only 5.1\% of the problems have C programs, and it is 31.5\% for PHP programs. Random splitting at the problem level can exacerbate this problem, resulting in very small test/validation dataset for C and PHP. Moreover, only a small number of problems have programs in all the 7 languages. It is beneficial to use these problems for evaluation, as they can provide a fair comparison across all the languages. Satisfying these two constraints, we take the following steps to create the data split:
\begin{enumerate}[leftmargin=*]
\item Out of all the problems that have programs (solutions) in all 7 languages, we randomly sample the test and validation sets for C. We start out with creating the splits for C in particular since it represents the smallest proportion of the dataset. 
\item Next, we first remove all the problems that have C programs. Out of all the problems that have programs in the remaining 6 languages (excluding C), we randomly sample the partial test and validation sets for PHP, so that the combined problems from this step and the previous one can be used as final test and validation sets of the PHP programs. 
\item Finally, we remove all the problems that have C programs or PHP programs. Since the remaining 5 languages have approximately same number of programs, we randomly sample the partial test and validation sets and use them for all the 5 languages. The final test and validation set for each of the 5 languages is the combination of these problems and problems from the previous two steps. This allows us to maintain a split ratio of approximately 85-5-10 (train-val-test) for all the 7 languages.
\end{enumerate}

Our splitting strategy provides a balanced split across languages and ensures there is no overlap between any evaluation set(test or validation) and any training set across all languages.

\subsection{More Details about XL Code Search}
For this task, we create 7 different datasets, one for each language where the chosen language is the query language and all other languages form the candidate solutions. For example, let us consider the dataset for C++ which contains entries like "1057-C++-1/1057-C\#-1". This basically represents the datapoint where the first snippet of problem ID 1057 in C++ is the query and the corresponding answer snippet is in C\#. However, this is not the only correct pairing, the dataset contains all the possible correct pairings which include \{1057-C\#-1, 1057-C-1, 1057-Python-1, 1057-Javascript-1, 1057-PHP-1, 1057-Java-1\}. When any of these solutions are present, the output candidate list is considered as a correctly chosen candidate. It should also be noted that all queries do not have exhaustively all other languages as candidate solutions.

\subsection{More Details about Evaluation Metrics}
\begin{itemize}[leftmargin=*]
    \item \textbf{BLEU:} Given an input code sample, we use BLEU \citep{papineni2002bleu} score to evaluate the $n$-gram overlap between the generated and the ground-truth target text and code.
    \item \textbf{CodeBLEU:} CodeBLEU \citep{ren2020codebleu} is designed for automatic evaluation of code synthesis. Besides $n$-gram match (as in BLEU), it also evaluates the code syntax via abstract syntax trees (AST) and code semantics via data-flow. We use CodeBLEU for code generation tasks like Code Translation and Code Synthesis. The original CodeBLEU does not support C and C++. We extend the CodeBLEU code to include these two languages. The related code is included in the GitHub repo.
    \item \textbf{Mean Reciprocal Rank (MRR):}
    The reciprocal rank is defined as the inverse of the rank of the first correct candidate for a given query. MRR is the mean of the reciprocal rank for all the queries in the test set. In order to evaluate our XL Code Search task, we modified the traditional definition of the \textbf{MRR} metric to account for the possibility of multiple correct candidate solutions. We modify it in the following manner:

    Given a query $q_i$ from the set of queries $Q = \{q_1, q_2 \ldots q_m\}$, candidate set $C_i = \{c_{i1}, c_{i2}, \ldots, c_{in}\}$ corresponding to $q_i$ and the answer set $A_i = \{a_{i1}, a_{i2}, \ldots, a_{ik}\}$ where $k \in [1,6]$, $r_{ij}$ is the reciprocal rank of the $j^{th}$ candidate $c_{ij}$ for $c_{ij} \in A_i$.
    
    $$ MRR_{q_{i}} = \frac{1}{k}\sum_{\substack{j=1 \\ c_{ij}\in A_i}}^{n} r_{ij}$$

    $$MRR_Q = \frac{1}{m}\sum_{i = 1}^{m} MRR_{q_{i}}$$
    
    
\end{itemize}

\begin{table*}[h]
\caption{\label{tab:data_sample_check_3}Example of the parallel alignment of the code data in four languages. The programs given here checks if a given number is divisible by 3 or not.}
\begin{tabular}{ |p{3cm}|p{3.2cm}|p{3cm}|p{3cm}| }
  \hline
\centering{\rule{0pt}{2ex}\textbf{C++}} & \centering{\textbf{Java}} & \centering{\textbf{\small{Python}}} & \qquad\quad\, \textbf{PHP}\\
\hline

\begin{lstlisting}[style=C++]
/* C++ program to find if a number is divisible by 3 or not */

#include<bits/stdc++.h>
using namespace std;
\end{lstlisting} &
\begin{lstlisting}[style=java]
/* Java program to find if a number is divisible by 3 or not */

class IsDivisible
{
\end{lstlisting}&
\begin{lstlisting}[style=py]
''' Python program to find if a number is divisible by 3 or not '''
\end{lstlisting}
&
\begin{lstlisting}[style=PHP]
/* PHP program to find if a number is divisible by 3 or not */

<?php
\end{lstlisting}\\
\hline
\begin{lstlisting}[style=C++]
/* Function to find that number divisible by 3 or not */

int check(string str)
{

\end{lstlisting} &
\begin{lstlisting}[style=java]
/* Function to find that number divisible by 3 or not */

static boolean check(String str)
{

\end{lstlisting}&
\begin{lstlisting}[style=py]
''' Function to find that number divisible by 3 or not '''

def check(num) :
\end{lstlisting}
&
\begin{lstlisting}[style=PHP]
/* Function to find that number divisible by 3 or not */

function check($str)
{

\end{lstlisting}\\
\hline
\begin{lstlisting}[style=C++]
/* Compute sum of digits */

int n = str.length();
int digitSum = 0;
for (int i=0; i<n; i++)
    digitSum += (str[i]-'0');
\end{lstlisting} &
\begin{lstlisting}[style=java]
/* Compute sum of digits */

int n = str.length();
int digitSum = 0;
for (int i=0; i<n; i++)
    digitSum += (str.charAt(i)-'0');
\end{lstlisting}&
\begin{lstlisting}[style=py]
''' Compute sum of digits '''

    digitSum = 0
    while num > 0 :
        rem = num % 10
        digitSum = digitSum + rem
        num = num / 10
\end{lstlisting}
&
\begin{lstlisting}[style=PHP]
/* Compute sum of digits */

$n = strlen($str);
$digitSum = 0;
for ($i = 0; $i < $n; $i++)
    $digitSum += ($str[$i] - '0');
\end{lstlisting}\\
\hline

\begin{lstlisting}[style=C++]
/* Check if sum of digits is divisible by 3 */

return (digitSum % 3 == 0);
}
\end{lstlisting} &
\begin{lstlisting}[style=java]
/* Check if sum of digits is divisible by 3 */

return (digitSum % 3 == 0);
}
\end{lstlisting}&
\begin{lstlisting}[style=py]
''' Check if sum of digits is divisible by 3 '''

    return (digitSum % 3 == 0)
\end{lstlisting}
&
\begin{lstlisting}[style=PHP]
/* Check if sum of digits is divisible by 3 */

return ($digitSum % 3 == 0);
}
\end{lstlisting}\\
\hline
\begin{lstlisting}[style=C++]
/* Driver code */

int main()
{
string str = ""1332"";
check(str)?  cout << ""Yes"" : cout << ""No "";
return 0;
}
\end{lstlisting} &
\begin{lstlisting}[style=java]
/* main function */

public static void main (String[] args)
{
String str = ""1332"";
if(check(str))
    System.out.println(""Yes"");
else
    System.out.println(""No"");
}
}
\end{lstlisting}&
\begin{lstlisting}[style=py]
''' main function '''

num = 1332
if(check(num)) :
    print ""Yes""
else :
    print ""No""
\end{lstlisting}
&
\begin{lstlisting}[style=PHP]
/* Driver code */

$str = "1332";
$x = check($str) ? "Yes" : "No ";
echo($x);
?>
\end{lstlisting}\\
\hline

\end{tabular}
\end{table*}

\begin{table*}[h]

\caption{\label{tab:data_sample_add_matrix}Example of the parallel alignment of the code data in four languages. The programs given here aim to find the LCM of two given numbers}

\begin{tabular}{ |p{3.2cm}|p{3cm}|p{3cm}|p{3cm}| }
  \hline
\centering{\rule{0pt}{2ex} \textbf{C\#}} & \centering{\textbf{JavaScript}} & \centering{\textbf{PHP}} & \qquad\quad\,\,\, \textbf{C}\\
\hline

\begin{lstlisting}[style=Csharp]
/* C# program to find LCM of two numbers */

using System;
class GFG {

\end{lstlisting} &
\begin{lstlisting}[style=java]
/* Javascript program to find LCM of two numbers */
\end{lstlisting}&
\begin{lstlisting}[style=PHP]
/* PHP program to find LCM of two numbers */

<?php
\end{lstlisting}
&
\begin{lstlisting}[style=C]
/*C program to find LCM of two numbers*/

#include <stdio.h>
\end{lstlisting}\\
\hline
\begin{lstlisting}[style=Csharp]
/* Recursive method to return gcd of a and b */

static int gcd(int a, int b)
{
    if (a == 0)
        return b;
    return gcd(b % a, a);
}
\end{lstlisting} &
\begin{lstlisting}[style=java]
/* Recursive function to return gcd of a and b */

function gcd(a, b)
{
if (b == 0)
    return a;
return gcd(b, a % b);
}
\end{lstlisting}&
\begin{lstlisting}[style=PHP]
/* Recursive function to return gcd of a and b */

function gcd( $a, $b)
{
   if ($a == 0)
        return $b;
    return gcd($b % $a, $a);
}
\end{lstlisting}
&
\begin{lstlisting}[style=C]
/* Recursive function to return gcd of a and b */

int gcd(int a, int b)
{
    if (a == 0)
        return b;
    return gcd(b % a, a);
}

\end{lstlisting}\\
\hline

\begin{lstlisting}[style=Csharp]
/* method to return LCM of two numbers */

static int lcm(int a, int b)
{
    return (a / gcd(a, b)) * b;
}

\end{lstlisting} &
\begin{lstlisting}[style=java]
/* Function to return LCM of two numbers */

function lcm(a, b)
{
    return (a / gcd(a, b)) * b;
}
\end{lstlisting}&
\begin{lstlisting}[style=PHP]
/* Function to return LCM of two numbers */

function lcm( $a, $b)
{
    return ($a / gcd($a, $b)) * $b;
}

\end{lstlisting}
&
\begin{lstlisting}[style=C]
/* Function to return LCM of two numbers */

int lcm(int a, int b)
{
    return (a / gcd(a, b)) * b;
}

\end{lstlisting}\\ \hline
\begin{lstlisting}[style=Csharp]
/* Driver method */

public static void Main()
{
int a = 15, b = 20;
Console.WriteLine("LCM of " + a + " and " + b + " is " + lcm(a, b));
}
}
\end{lstlisting} &
\begin{lstlisting}[style=java]
/* Driver program to test above function */

let a = 15, b = 20;
document.write("LCM of " + a + " and " + b + " is " + lcm(a, b));
\end{lstlisting}&
\begin{lstlisting}[style=PHP]
/* Driver Code */

$a = 15;
$b = 20;
echo "LCM of ",$a, " and " ,$b, " is ", lcm($a, $b);
?>
\end{lstlisting}
&
\begin{lstlisting}[style=C]
/* Driver program to test above function */

int main()
{
int a = 15, b = 20;
printf("LCM of %d and %d is %d ", a, b, lcm(a, b));
return 0;
}
\end{lstlisting}\\

\hline
\end{tabular}
\end{table*}

\begin{table*}[h]
\caption{\label{tab:data_sample_array_rotation} Example of the parallel alignment of the code data in all seven languages. The Programs given here aim to find two elements whose sum is closest to zero.}\vspace{0.05in}
\begin{tabular}{ |p{3.2cm}|p{3cm}|p{3cm}|p{3cm}| }
  \hline
\centering{\rule{0pt}{2ex}\textbf{C++}} & \centering{\textbf{Java}} & \centering{\textbf{\small{Python}}} & \qquad\quad\,\, \textbf{C\#}\\
\hline
\begin{lstlisting}[style=C++]
/* C++ code to find Two elements whose sum is closest to zero */

# include <bits/stdc++.h>
# include <stdlib.h>
# include <math.h>
using namespace std;
void minAbsSumPair(int arr[], int arr_size)
{
int inv_count = 0;
int l, r, min_sum, sum, min_l, min_r;
\end{lstlisting} &
\begin{lstlisting}[style=java]
/* Java code to find Two elements whose sum is closest to zero */

import java.util.*;
import java.lang.*;
class Main
{
static void minAbsSumPair(int arr[], int arr_size)
{
int inv_count = 0;
int l, r, min_sum, sum, min_l, min_r;
      
\end{lstlisting}&
\begin{lstlisting}[style=py]
 ''' Python3 code to find Two elements whose sum is closest to zero '''

def minAbsSumPair(arr,arr_size):
    inv_count = 0

\end{lstlisting}
&
\begin{lstlisting}[style=Csharp]
/* C# code to find Two elements whose sum is closest to zero */

using System;
class GFG
{
static void minAbsSumPair(int []arr, int arr_size)
{
int l, r, min_sum, sum, min_l, min_r;
    
\end{lstlisting}\\[1em]
\hline

\begin{lstlisting}[style=C++]
/* Array should have at least two elements */

if (arr_size < 2)
{
    Console.Write("Invalid Input");
    return;
}
\end{lstlisting} &
\begin{lstlisting}[style=java]
/* Array should have at least two elements */

if(arr_size < 2)
{
    document.write("Invalid Input");
    return;
}
    
      
\end{lstlisting}&
\begin{lstlisting}[style=py]
 ''' Array should have at least two elements '''

    if arr_size < 2:
        print("Invalid Input")
        return


\end{lstlisting}
&
\begin{lstlisting}[style=Csharp]
/* Array should have at least two elements */

if (arr_size < 2)
{
    Console.Write("Invalid Input");
    return;
}
    
\end{lstlisting}\\[1em]
\hline

\begin{lstlisting}[style=C++]
/* Initialization of values */

min_l = 0;
min_r = 1;
min_sum = arr[0] + arr[1];
for(l = 0; l < arr_size - 1; l++)
{
    for(r = l + 1; r < arr_size; r++)
    {
    sum = arr[l] + arr[r];
    if(abs(min_sum) > abs(sum))
    {
        min_sum = sum;
        min_l = l;
        min_r = r;
    }}}}

\end{lstlisting} &
\begin{lstlisting}[style=java]
/* Initialization of values */

min_l = 0;
min_r = 1;
min_sum = arr[0] + arr[1];
for(l = 0; l < arr_size - 1; l++)
{
for(r = l+1; r < arr_size; r++)
{
  sum = arr[l] + arr[r];
  if(Math.abs(min_sum) > Math.abs(sum))
  {
    min_sum = sum;
    min_l = l;
    min_r = r;
  }}}}

\end{lstlisting}&
\begin{lstlisting}[style=py]
 ''' Initialization of values '''

    min_l = 0
    min_r = 1
    min_sum = arr[0] + arr[1]
    for l in range (0, arr_size - 1):
        for r in range (l + 1, arr_size):
            sum = arr[l] + arr[r]                
            if abs(min_sum) > abs(sum):        
                min_sum = sum
                min_l = l
                min_r = r
\end{lstlisting}
&
\begin{lstlisting}[style=Csharp]
/* Initialization of values */

min_l = 0;
min_r = 1;
min_sum = arr[0] + arr[1];
for (l = 0; l < arr_size - 1; l++)
{
    for (r = l+1; r < arr_size; r++)
    {
        sum = arr[l] + arr[r];
        if (Math.Abs(min_sum) > Math.Abs(sum))
        {
            min_sum = sum;
            min_l = l;
            min_r = r;
        }}}}
\end{lstlisting}\\[1em]
\hline

\begin{lstlisting}[style=C++]
/* Driver Code */

int main()
{
    int arr[] = {1, 60, -10, 70, -80, 85};
    minAbsSumPair(arr, 6);
    return 0;
}
\end{lstlisting} &
\begin{lstlisting}[style=java]
/* main function */

public static void main (String[] args)
{
    int arr[] = {1, 60, -10, 70, -80, 85};
    minAbsSumPair(arr, 6);
}
}
\end{lstlisting}&
\begin{lstlisting}[style=py]
 ''' Driver program to test above function '''

arr = [1, 60, -10, 70, -80, 85]
minAbsSumPair(arr, 6);
\end{lstlisting}
&
\begin{lstlisting}[style=Csharp]
/* main function */

public static void Main ()
{
    int []arr = {1, 60, -10, 70, -80, 85};
    minAbsSumPair(arr, 6);
}
}
\end{lstlisting}\\[1em]
\hline

\end{tabular}

\end{table*}

\begin{table*}[h]
\begin{tabular}{ |p{3.2cm}|p{3cm}|p{3cm}|p{3cm}| }
\hline
\centering{\rule{0pt}{2ex}\textbf{JavaScript}} & \centering{\textbf{PHP}} & \centering{\textbf{C}}&  \\
\hline
\begin{lstlisting}[style=java]
/* JavaScript code to find Two elements whose sum is closest to zero */

function minAbsSumPair( arr,  arr_size)
{
var inv_count = 0;
var l, r, min_sum, sum, min_l, min_r;
\end{lstlisting} &
\begin{lstlisting}[style=PHP]
/* PHP program to find the Two elements whose sum is closest to zero */

function minAbsSumPair($arr, $arr_size)
{
$inv_count = 0;
\end{lstlisting}&
\begin{lstlisting}[style=C++]
/* C code to find Two elements whose sum is closest to zero */

# include <stdio.h>
# include <stdlib.h>
# include <math.h>
void minAbsSumPair(int arr[], int arr_size)
{
int inv_count = 0;
int l, r, min_sum, sum, min_l, min_r;
\end{lstlisting}
&\\
\hline
\begin{lstlisting}[style=java]
/* Array should have at least two elements */

if(arr_size < 2)
{
    document.write("Invalid Input");
    return;
}
    
\end{lstlisting} &
\begin{lstlisting}[style=PHP]
/* Array should have at least two elements */

if($arr_size < 2)
{
    echo "Invalid Input";
    return;
}
    
\end{lstlisting}&
\begin{lstlisting}[style=C++]
/* Array should have at least two elements */

if(arr_size < 2)
{
printf("Invalid Input");
return;
}
  
\end{lstlisting}
&\\
\hline
\begin{lstlisting}[style=java]
/* Initialization of values */

min_l = 0;
min_r = 1;
min_sum = arr[0] + arr[1];
for(l = 0; l < arr_size - 1; l++)
{
    for(r = l + 1; r < arr_size; r++)
    {
        sum = arr[l] + arr[r];
        if(Math.abs(min_sum) > Math.abs(sum))
    {
        min_sum = sum;
        min_l = l;
        min_r = r;
    }}}}
\end{lstlisting} &
\begin{lstlisting}[style=PHP]
/* Initialization of values */

$min_l = 0;
$min_r = 1;
$min_sum = $arr[0] + $arr[1];
for($l = 0; $l < $arr_size - 1; $l++)
{
    for($r = $l+1; $r < $arr_size; $r++)
    {
    $sum = $arr[$l] + $arr[$r];
    if(abs($min_sum) > abs($sum))
    {
        $min_sum = $sum;
        $min_l = $l;
        $min_r = $r;
    }}}}
\end{lstlisting}&
\begin{lstlisting}[style=C++]
/* Initialization of values */

min_l = 0;
min_r = 1;
min_sum = arr[0] + arr[1];
for(l = 0; l < arr_size - 1; l++)
{
for(r = l+1; r < arr_size; r++)
{
  sum = arr[l] + arr[r];
  if(abs(min_sum) > abs(sum))
  {
    min_sum = sum;
    min_l = l;
    min_r = r;
  }}}}
\end{lstlisting}
&\\
\hline
\begin{lstlisting}[style=java]
/* Driver Code */

arr = new Array(1, 60, -10, 70, -80, 85);
minAbsSumPair(arr, 6);
\end{lstlisting} &
\begin{lstlisting}[style=PHP]
/* Driver Code */

$arr = array(1, 60, -10, 70, -80, 85);
minAbsSumPair($arr, 6);

?>
\end{lstlisting}&
\begin{lstlisting}[style=C++]
/* Driver program to test above function */

int main()
{
  int arr[] = {1, 60, -10, 70, -80, 85};
  minAbsSumPair(arr, 6);
  getchar();
  return 0;
}
\end{lstlisting}
&\\
\hline
\end{tabular}

\end{table*}

\section{Dataset Information}
The dataset and the code used in this paper can be found at \url{https://github.com/reddy-lab-code-research/XLCoST}. Due to the large size of the data, it is shared through a Google Drive link (provided in the repository). The dataset and the code are distributed under CC BY-SA License 4.0  and Apache License 2.0, respectively.

\subsection{Motivation}
As described in the main paper, the primary motivation behind the creation and release of this data is to potentially facilitate and foster research in the domain of Deep Learning for Software Engineering. Code related tasks have garnered a lot of attention by the community in the past few years but it has been our observation that the availability of high quality, parallel data across multiple languages which is required to be able to produce advances in this domain, is still limited. We discuss in the main paper as well, how most of the widely used datasets are either limited to just a few language pairs or are limited in size. With the release of this dataset we aim to fill both of those gaps,
and to give the research community better tools in order to solve code-related tasks.

\subsection{Intended Use}
The primary intended uses of the dataset are to encourage development and validation of models/methods for code related tasks such as translation, summarization, synthesis, and search. The link to the dataset can be found in the README of the GitHub repository. Code required to reproduce results and baseline scores can also be found in the GitHub README file. Readers will need to cite the original dataset when using it in their experiments or making modifications to it.

\subsection{Author Statement}
The IP policies and regulations for GeeksForGeeks were carefully followed and we confirm that no data privacy policy was violated when collecting the data. We bear all responsibility in case of violation of rights. We confirm that the dataset is distributed under CC BY-SA License 4.0.

\subsection{Maintenance}
The dataset will be actively maintained by the authors. Issues can be reported via raising an issue on GitHub or e-mail to one of the authors. The dataset will be hosted on Google Drive since its large size is not supported by GitHub. Any changes to hosting will be reflected in the links on the GitHub repository. The authors may also update the dataset by adding more datapoints, or in case issues are reported by other parties or are found by the authors themselves. Any such updates to the data will be documented on GitHub. 

\subsection{Societal Impact}
Since deep learning models have become larger, the amount of computational power needed to train and maintain them has also increased. An unintended consequence of this has been the increased carbon footprint of deep learning research as a result of running large number of experiments to validate hypotheses. As our dataset aims to facilitate further research in the domain, it would also end up having this societal impact, albeit indirectly so. We would encourage users to use compute and memory efficient methods when carrying their research using this dataset.

\end{document}